# Leadership et changements institutionnels[*]

Matata Ponyo Mapon[†]

Jean-Paul K. Tsasa[‡]

**Résumé**. Des changements institutionnels à grande échelle tels que la réforme de l'administration publique, la lutte systémique contre la corruption et les rentes ou la promotion des droits civils requièrent l'adhésion et l'implication de plusieurs participants ou forces. Nous utilisons le cadre standard de la *théorie des jeux coopératifs* développé par Ichiishi (1983, pp. 78-149) et le réaménageons pour: (i) établir analytiquement une différence entre dirigeant politique et leader politique; (ii) étudier formellement les interactions bidirectionnelles entre un dirigeant politique et ses partisans; (iii) examiner le rôle du leadership dans la conduite des réformes structurelles. Nous montrons qu'un dirigeant politique peut être à la fois partisan et non-partisan, alors qu'un leader politique ne peut qu'être non-partisan. Suivant cette distinction, nous dérivons la probabilité de réussite d'un changement institutionnel, ainsi que la nature du gain que générerait un tel changement sur la population bénéficiaire. En nous basant sur les restrictions de ce modèle mathématique simple, nous montrons à l'aide des quelques exemples pertinents, tirés de l'expérience congolaise entre 2012 et 2016, que les changements institutionnels peuvent effectivement profiter à la majorité de la population, lorsque les dirigeants politiques sont véritablement « partisans », i.e. lorsqu'ils font preuve d'un leadership efficient dans la conduite des réformes institutionnelles.

**Mots-clés**: Leadership, réformes structurelles, interactions partisanes.
**Classification JEL**: C71, D02, O43, P41.

**Abstract** (*Title: Leadership and Institutional Reforms*) Large-scale institutional changes require strong commitment and involvement of all stakeholders. We use the standard framework of cooperative game theory developed by Ichiishi (1983, pp. 78-149) to: (i) establish analytically the difference between policy maker and political leader; (ii) formally study interactions between a policy maker and his followers; (iii) examine the role of leadership in the implementation of structural reforms. We show that a policy maker can be both partisan and non-partisan, while a political leader can only be non-partisan. Following this distinction, we derive the probability of success of an institutional change, as well as the nature of the gain that such a change would generate on the beneficiary population. Based on the restrictions of this simple mathematical model and using some evidence from the Congolese experience between 2012 and 2016, we show that institutional changes can indeed benefit the majority of the population, when policy makers are truly "partisan".

---

[†] Professeur, Faculté d'Administration des affaires et Sciences Économiques. Université Protestante au Congo, Kinshasa, RDC. Email: matataponyo@congochallenge.cd.
[‡] PhD candidate, Département des sciences économiques. Université du Québec, Montréal, CANADA, Bureau: DS-5820. Fax: +1 514 987-8494. E-mail: tsasa.jean-paul@courrier.uqam.ca.

## 1. Introduction

> *Les dirigeants politiques ont le pouvoir de désinformer et de manipuler, mais ils ont également le pouvoir d'inspirer le changement.* (Levi 2006, p. 12)

La littérature économique sur les changements institutionnels reconnait l'importance des leaders (et du leadership) sur le profil du développement économique d'un pays. Cette importance peut s'expliquer par le fait que, dans de nombreuses situations, les dirigeants politiques sont au cœur des réformes structurelles, et donc au cœur de la transformation espérée des institutions politiques et économiques d'un pays[1]. Autrement dit, le succès ou l'échec dans la mise en œuvre des réformes structurelles repose fondamentalement sur l'efficacité des dirigeants politiques qui sont élus ou nommés dans les différents postes de responsabilité[2]. À son tour, de quoi dépend l'efficacité d'un dirigeant politique dans le processus de mise en œuvre des réformes structurelles ou des changements institutionnels à grande échelle? Dans ce papier, nous partons de l'idée que l'efficacité d'un dirigeant politique dépend essentiellement des quatre facteurs-clés.

Premièrement, de son habileté à identifier, dans des circonstances et moments appropriés, les opportunités de changement (Nohria and Khurana 2010, p. 29, Metcalf et Benn 2013, p. 372; Barling 2014, p. 9). En effet, les réformes structurelles requièrent généralement une coordination des croyances et des actions d'un grand nombre d'individus au sein de différents groupes et couches de la population. Ces derniers ont des priorités ou préférences souvent différentes et même divergentes. Certains préfèrent le changement, d'autres en revanche préfèrent le statu quo. Dans ce contexte, les interactions stratégiques entre un dirigeant politique et les différents groupes concernés devraient s'articuler autour de la nécessité d'ajuster conséquemment l'offre et la demande des réformes. Le sens de cet ajustement dépendra de la nature intrinsèque des dirigeants politiques selon qu'ils sont non-partisans (c'est-à-dire « leader ») ou partisans.

---

[1] Par exemple, le Botswana a connu le taux de croissance du PIB par habitant plus élevé que tout autre pays du monde entre 1965 et 1998. Selon Acemoglu et al. (2003), cela s'est produit grâce à l'existence des institutions précoloniales inclusives, imposant des contraintes aux élites politiques, mais aussi à cause d'un certain nombre de décisions critiques prises par les dirigeants de l'après-indépendance, en particulier les présidents Seretse Khama et Quett Masire. Une analyse similaire est proposée dans Frankel (2016) pour l'île Maurice, où l'auteur souligne le leadership de Sir Seewoosagur Ramgoolam dans le choix du type des institutions promues dès 1968 au moment de l'indépendance de l'île. Voir aussi Hermalin (1998) et Alesina et al. (2006).

[2] Il est important de souligner ici que tout dirigeant politique n'est pas automatiquement un « leader ». La figure 1 en propose une illustration simplifiée. Dans ce papier, nous conférons au terme « leader » une signification conservatrice. En effet, un leader est celui qui a le pouvoir d'inspirer les changements (cf. Levi 2006, p. 12), mais aussi de les imprimer effectivement au gré des groupes de pression (souvent, une minorité) ou obstacles divers. De ce fait, le leadership peut donc être analytiquement perçu comme la probabilité qu'un dirigeant politique soit susceptible d'inspirer et de provoquer effectivement un changement en faveur de la majorité de la population. Une discussion documentée à ce sujet peut être trouvée dans Matata et Tsasa (2019, 2020).

Figure 1: Illustration simplifié d'un espace politique

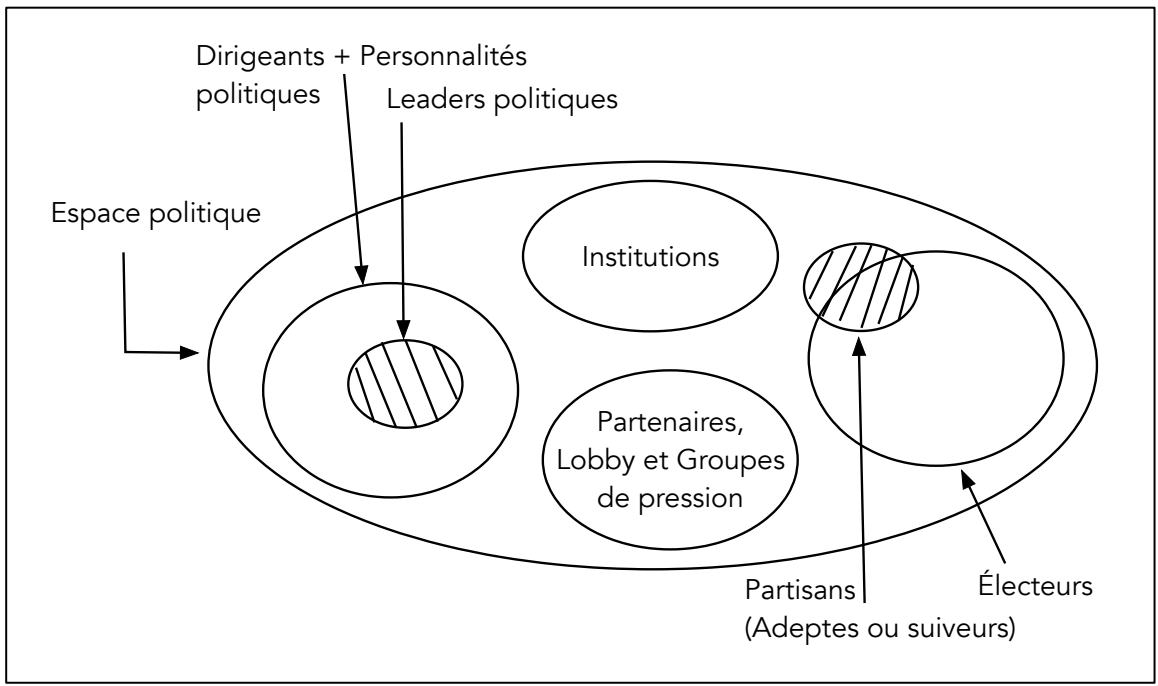

Note: La figure 1 présente une illustration simplifiée d'un espace politique. Par exemple, elle précise que tout dirigeant politique n'est pas nécessairement un leader et que tout partisan d'un dirigeant politique n'est pas forcement électeur. Source: Figure conçue et élaborée par les auteurs.

Deuxièmement, nous soutenons que l'efficacité d'un dirigeant politique dépend de sa capacité à attirer et à conserver un groupe des collaborateurs et des partisans engagés. Un tel groupe œuvre généralement avec une vigueur extraordinaire et souvent en se sacrifiant. Ainsi, au sens de Willner (1985), un groupe de partisans engagés constitue, en réalité, un instrument malléable que les dirigeants politiques peuvent utiliser à volonté.

Troisièmement, l'efficacité d'un dirigeant politique dans le processus de mise en œuvre des réformes structurelles dépend de son aptitude à mobiliser la participation des partisans (adeptes, militants ou « suiveurs »), mais aussi des non-partisans, autrement dit de son habileté à communiquer efficacement avec la population. En effet, peu importe le contexte ou les motivations, s'engager à soutenir un mouvement ou à suivre un dirigeant politique est généralement très couteux. De ce fait, les individus ne deviennent des partisans engagés que s'ils s'attendent à des gains, ceux-ci peuvent être d'ordre matériel, financier, professionnel ou immatériel (par exemple, un idéal, une idéologie ou une conviction). Cela crée donc une offre et une demande d'interactions partisanes, appelées aussi dans certains contextes offre et demande de leadership[3].

---

[3] Notamment lorsque les partisans font face à un leader (cf. Figure 2 dans la section 3). En parallèle, notons ici que l'économie politique des réformes insiste sur les interactions stratégiques entre différents groupes d'intérêt dans l'explication du succès ou de l'échec d'une réforme structurelle. Suivant cette perspective, le succès ou l'échec d'une réforme dépend de la sensibilité des groupes sociaux à l'incertitude. En effet, pour certains groupes, les bénéfices nets attendus des réformes,

Enfin, quatrièmement, la perception qu'a le public de la capacité d'un dirigeant politique à initier et à mener avec succès les réformes joue également un rôle critique dans la détermination de l'efficacité d'un dirigeant politique à provoquer effectivement le changement.

La mise ensemble de ces quatre facteurs soulève, donc naturellement, les questions traditionnelles d'interactions stratégiques, de sélection, de coordination et de croyance[4]. En conséquence, le cadre approprié pour examiner ces questions est vraisemblablement la théorie des jeux (voir, par exemple, Ichiishi (1983)). Nous utilisons ce cadre d'analyse, notamment, pour montrer que les problèmes que posent la coordination, la sélection, les interactions stratégiques et la croyance imposent généralement des restrictions qui permettent de dissocier analytiquement un dirigeant politique et un leader politique, et donc de déterminer suivant cette distinction les gains escomptés d'un changement institutionnel sur la population (minorité *versus* majorité).

Nous nous servons du cas de la République démocratique du Congo (RDC) comme cadre de référence, à l'effet d'illustrer les interactions entre leadership et réformes structurelles. La RDC est un pays riche en ressources naturelles, mais économiquement pauvres[5]. Inscrire une telle économie dans un sentier de croissance stable à long terme requiert préalablement et fondamentalement la mise en œuvre des réformes structurelles de grandes envergures, susceptibles d'accroître l'efficacité économique et de promouvoir les institutions inclusives. La mise en œuvre des réformes structurelles dans des pays tels que la RDC, ou en général dans les pays en développement, soulève de nombreux enjeux de natures diverses. Au centre de ces enjeux, au moins trois aspects institutionnels majeurs sont en présence.

Le premier aspect est que la réforme peut se retrouver bloquée ou aboutir à des résultats incertains. Cela en raison de nombreux facteurs sociopolitiques parmi lesquels, des comportements opportunistes des politiciens au demeurant idéologiquement polarisés, des conflits redistributifs, ou des conflits au sein des partis de gouvernement, ou enfin mais sans être exhaustif, de l'hétérogénéité des intérêts des groupes de pression en présence. Ainsi, par exemple au plus fort des programmes d'ajustement structurel au début des

---

peuvent aller de pair avec une incertitude sur les bénéfices nets différentiels que d'autres groupes concurrents pourraient retirer. Dans ces conditions, les premiers sont de nature à ne pas coopérer avec le gouvernement (Alesina et Drazen, 1991). Dans d'autres circonstances, certains types de réformes induisent un aléa des bénéfices nets attendus chez certains groupes, ce qui conduit à « un statu quo généralisé », du fait de l'incertitude des effets redistributifs des réformes (cf. Rodrik 1996).

[4] Par exemple, lorsque les croyances au sein du groupe censé bénéficier des réformes structurelles sont pessimistes, l'avènement effectif du changement serait plus difficile, du fait notamment de la faible partition et implication de la population cible.

[5] L'IDH de la RDC pour 2018 s'établit à 0,459, ce qui place le pays dans la catégorie « développement humain faible » et au 179e rang parmi 189 pays et territoires (cf. PNUD 2019).

années 1990, le Fonds monétaire international constatait-il que « les réformes économiques demeurent incomplètes, et la viabilité des finances publiques est à peine allusive »[6].

Le deuxième aspect porte sur la définition des institutions de développement. Pour Acemoglu et Robinson (2013), le point de départ est que ce sont les institutions politiques qui déterminent les institutions économiques. On peut en effet distinguer quatre types d'institutions: (i) les institutions économiques extractives, caractérisées par l'absence d'ordre, de règles et de lois garantissant la sécurité des droits de propriété, puis dominées par des barrières à l'entrée et des réglementations qui contrarient le jeu libre des marchés; (ii) les institutions politiques extractives, marquées par la détention du pouvoir par un petit nombre d'individus, en l'absence d'équilibre des pouvoirs, de contre-pouvoirs et de contrôle de l'exécution du pouvoir par la force de loi. À l'extrême, elles relèvent de « l'absolutisme politique »; (iii) les institutions économiques inclusives, marquées par la défense des droits de propriété et la sécurité des contrats, elles reposent sur l'ordre, le fonctionnement libre des marchés, la libre entrée et sortie, avec un soutien de l'État aux marchés via des services publics de qualité, un accès à l'éducation et aux opportunités économiques pour la grande majorité de la population; (iv) les institutions politiques inclusives, déterminées par le pluralisme politique, où les dirigeants sont soumis à des contrôles, l'exercice du pouvoir soumis à des règles de droit et accompagné de contre-pouvoirs. En sus, ces institutions peuvent être formelles (gouvernement, parlement, partis politiques officiels, administration centrale, etc.) ou informelles (notamment les diverses émanations des organisations religieuses, les ethnies, les clans et leurs centres de pouvoirs, les réseaux ésotériques et diverses fraternités, les forces politiques souterraines ou en lien avec les forces de sécurité, etc.).

Enfin, le troisième aspect relève du rôle qu'exerce le leadership en matière de succès des réformes en vue du développement. En Afrique, les États se soumettent rarement au droit, et les individus eux-mêmes coopèrent rarement en vue du respect de l'État de droit, lequel ne s'impose ni aux classes dirigeantes, ni véritablement aux individus. De ce fait, les normes juridiques anciennes ou nouvelles, les institutions, anciennes ou nouvelles, selon les termes de Mwayila Tshiyembe (2015, p. 22), « roulent sur le corps social sans jamais le pénétrer ». Dans ce contexte, il y a donc nécessité d'un leader efficace pour imposer des réformes proactives, visant notamment la remise en ordre de l'État, de ses attributs, et la diffusion auprès du corps social de la dynamique de progrès.

Le reste du papier s'organise comme suit. La section 2 discute de la place du leadership dans la mise en œuvre des réformes structurelles. La section 3 présente le modèle d'offre et de demande du leadership. La section 4 discute brièvement des implications analytiques du modèle d'analyse. La section 5 procède à un diagnostic détaillé du lien entre leadership et conduite des réformes institutionnelles en RDC durant la période 2012-2016. La section 6 conclut.

---

[6] Cf. FMI (1993, p. 40).

## 2. Leadership et mise en œuvre des réformes structurelles

D'entrée de jeu, précisons ce que nous entendons par le leadership. Comme cela a été largement discuté dans Matata Ponyo et Tsasa (2019, 2020), il n'existe pas une définition universelle du leadership. Aussi faut-il encore souligner, ici, que le fait d'occuper un poste d'autorité ne fait pas d'un dirigeant politique, d'un acteur social ou d'un gestionnaire un leader. Enfin, comme l'intensité d'une lumière, le leadership peut être médiocre (faible luminosité), excessif (luminosité exagérément élevée) ou efficient (luminosité optimale). En physique, la lumière peut être appréhendée ou caractérisée indépendamment des épithètes qu'on lui rattache. Ainsi nous proposons-nous de définir, ci-après, le leadership indépendamment de tout attribut ou épithète qu'on peut lui rattacher.

Dans ce papier, le leadership est compris comme étant la capacité pour un individu (i) d'avoir une vision claire du progrès de son organisation ou de son pays, (ii) de définir des objectifs globaux et intermédiaires précis et les mettre en œuvre, (iii) de résister aux pesanteurs de toute nature pouvant fragiliser ou anéantir la démarche entreprise et (iv) de vaincre, le cas échéant, les éventuels obstacles érigés par les groupes de pression et autres lobbyistes qui cherchent à maintenir le statu quo ou à préserver leurs intérêts privés au détriment de ceux de la majorité.

Dans la conduite des réformes structurelles de grande envergure, différents attributs peuvent être rattachés au leadership selon la nature des objectifs à réaliser, mais aussi l'ampleur des contraintes ou obstacles à surmonter. En effet, comme le fait valoir Goleman (2017), de nouvelles recherches suggèrent que les cadres les plus efficaces utilisent une collection de styles de leadership distincts, chacun dans la bonne mesure et au bon moment. En référence à la taxonomie proposée par l'auteur, au moins, six attributs peuvent être associés à un leader dont, ci-après, nous nous proposons de résumer parcimonieusement les avantages et les inconvénients.

Le leader *directif*. Il s'agit d'un leadership injonctif ou coercitif, illustrant la figure archaïque d'un chef très directif. Le leader directif laisse peu de place aux initiatives. Il impose les actions à mener, sans expliquer la vision globale. Il attend une exécution immédiate, et contrôle ce qui est fait. Il a tendance à manager ses équipes à l'échelle de tâches concrètes (micro-management) plutôt que de regarder la vision d'ensemble (macro-management). Il donne des ordres qui doivent être exécutés. L'avantage est qu'il a le contrôle, il dirige ses équipes et peut obtenir des avancées rapides et concrètes s'il est bien suivi. Ce leadership est optimal en cas de crise, quand il faut prendre un virage serré, ou pour serrer la vis aux collaborateurs qui posent problème, qui ne suivent pas les consignes ou qui attendent passivement qu'on leur donne des tâches bien précises. Le principal inconvénient est que s'il est mal utilisé, par exemple utilisé exclusivement et avec trop d'énergie, ce style de leadership engendre une résistance passive des collaborateurs. Aussi, par ailleurs, ce style de leadership ne fonctionne pas lorsqu'on est en face des tâches qui sont plus complexes, notamment celles qui font appel à l'initiative ou à la créativité. Souvent faut-il, en sus, souligner que ce style de leadership entraîne des effets négatifs sur le climat et la motivation intrinsèque des équipes, surtout lorsque ces dernières ne ressentent plus la confiance de leur chef et n'appréhendent plus le sens de leur travail.

Le leader *chef de file*. Un peu moins autoritaire que le style du leader directif, ce style de leadership n'en est pas moins exigeant. Le leader chef de file attend l'excellence. Il montre l'exemple d'un haut niveau de performance, toujours plus et mieux, et attend le même standard de ses équipes. Il donne le rythme, aux équipes de suivre. Il manque de patience avec les moins performants. Comme le leader directif, il est davantage centré sur les tâches que sur la vision d'ensemble (selon le slogan: Regardez-moi et faites comme moi). Sur le climat de l'équipe, l'effet de ce style est plutôt négatif dans l'ensemble, en ce sens que seuls ceux qui parviennent à suivre le leader peuvent garder toute leur motivation, alors que les autres risquent de se décourager, se démotiver avec l'impression de ne pas être à la hauteur. De ce fait, si le leader chef de file est suivi, il obtient des équipes les résultats qu'il attend. Par ailleurs, un autre inconvénient pour ce style de leadership est qu'il reprend souvent les tâches à son compte (seule garantie d'avoir exactement ce qu'il veut), d'où un surcroît de travail. Il ne favorise pas la montée en compétences des équipes.

Le leader *visionnaire*. Il s'agit ici d'un leader plus mobilisateur, charismatique qui séduit ses subordonnés. Le leader visionnaire fédère autour d'une vision. Il a l'art de communiquer une vision inspirante pour tous et pour chacun. Contrairement au leader directif et au leader chef de file, il privilégie une vue d'ensemble. Il explique à ses équipes le sens, la vision, le cap, et laisse faire. Il compte sur ses managers pour s'occuper du « comment » et de la mise en œuvre de sa vision. Ce type de leadership a des effets positifs sur le climat de l'équipe grâce notamment au charisme et à l'empathie du leader. Ce style est approprié pour montrer la voie et donner du sens aux changements dans l'État (éventuellement l'entreprise ou l'association). En revanche, ce style de leadership semble plus populiste qu'efficace en temps de crise, quand il faut agir vite et être plus directif. L'exigence, dans ce cas, serait donc de parvenir à faire passer la vision (être convaincant, inspirant, et pas trop stratosphérique) et d'avoir un management qui sait traduire cette vision en actes.

Le leader *collaboratif*. C'est un leader qui croit en l'harmonie, et cherche la cohésion. Il favorise les interactions (échanger, travailler ensemble) et comprend les besoins de l'équipe, qu'il cherche à satisfaire. Il est en mesure d'organiser des séminaires de team-building et à apaiser tous les conflits. Ce style de leadership a des effets positifs sur le climat de l'équipe car renforçant la motivation et la confiance des collaborateurs, au moins à court terme. Ce style de leadership renforce la cohésion d'équipe et donne à chacun les moyens de travailler dans les meilleures conditions. Il est approprié pour apaiser des tensions dans une équipe, soutenir la motivation en période difficile. Il permet de faire travailler ensemble une équipe d'experts qui habituellement travaillent séparément. En revanche, il s'avère trop doux pour les collaborateurs très performants qui attendent un modèle plus proche du chef de file, et ne permet pas aux individus de se sentir valorisés à titre individuel, car le leader collaboratif ne pense qu'en équipe.

Le leader *participatif*. Ce style de leadership cherche le consensus par la voie démocratique. Pacificateur et doté d'une bonne écoute, il appelle les idées de tous. C'est un convaincu de l'intelligence collective. Il sollicite volontiers les uns et les autres dans une attitude ouverte, de dialogue. Ce leader tranche rarement sans avoir au préalable écouté d'autres avis. Ce style de leadership améliore la créativité collective et l'innovation. Il permet de bénéficier d'une intelligence collective. Il est optimal pour obtenir l'unanimité ou l'engagement, ou pour recueillir des idées de la part de collaborateurs de valeur, notamment quand le leader est le nouveau manager d'une équipe qui en sait plus que lui-même.

Le leader *coach* investit sur les personnes. Il passe du temps avec elles et les aide à développer leurs forces et résoudre leurs faiblesses, en ligne avec leurs objectifs professionnels. Il cherche l'autonomie de chacun et la construction d'équipes compétentes. Il vise le long terme, tout en tenant compte des objectifs plus proches. Ce style est optimal pour aider les collaborateurs à améliorer leur productivité, à développer leurs ressources, être plus efficace dans l'autonomie. En revanche, il est difficile à mettre en œuvre, car il s'agit de guider tout en laissant l'autonomie. Des qualités d'écoute et de bienveillance sont requises, ainsi qu'une confiance dans les capacités de chacun à s'améliorer. Par ailleurs, ce style ne fonctionne pas avec les collaborateurs qui attendent une liste des tâches précises à exécuter.

En conséquence du développement qui précède, un leader efficient devrait naturellement adopter une approche éclectique consistant à tirer profit de meilleurs aspects de chaque style de leadership sus-évoqué, notamment: la capacité à attirer et à conserver un groupe des collaborateurs et des partisans engagés; l'aptitude à mobiliser la participation des partisans (adeptes, militants ou « suiveurs ») mais aussi des non-partisans, et donc à bien communiquer; l'habileté à identifier, dans des circonstances et moments appropriés, les opportunités de changement; l'intelligence de simuler la perception qu'ont les autres ou le public de la pertinence de ses actions. Le fait pour un réformateur de révéler les qualités d'un leader efficient ne garantit pas automatiquement la réussite d'une réforme structurelle à mettre en œuvre. En effet, la conduite des réformes structurelles de grande envergure, devant aboutir à des innovations et changements institutionnels majeurs, sont des entreprises difficiles à concrétiser, notamment dans le contexte de la RDC, un pays ayant connu des crises sécuritaires, tensions politiques et chocs macroéconomiques assez sévères en termes d'ampleur et assez significatifs en terme de longévité (conflits armés, instabilité politique, endettement extérieur explosif, hyperinflation, pillages, etc.). Quelles que soient les qualités du leader, l'économie politique des réformes et l'expérience contemporaine dans les pays développés comme dans les pays en développement, conduit à retenir quelques conditions supplémentaires à même de renforcer les chances de succès d'une réforme, dont on peut s'inspirer ici.

*Une bonne réforme doit être claire, lisible et transparente, les résultats devant être connus et affichés*. La dissémination des informations sur les objectifs de la réforme, les moyens pris pour la réaliser, et les résultats obtenus, est cruciale aussi bien pour le soutien à la réforme, et ex-post pour la crédibilité du réformateur.

*Pour qu'une réforme soit transparente, il convient qu'elle soit évaluée*. Les critères de suivi-évaluation sont absolument indispensables à la réussite d'une bonne réforme, car ils permettent de la réorienter le cas échéant ou de l'ajuster instantanément.

*La bonne réforme doit s'ajuster au cycle*. Les réformes impopulaires sont effectuées dans les pays développés en période de récession et rarement en période de croissance, alors que ce devrait être l'inverse. Une bonne réforme doit viser à rendre compatibles les objectifs macroéconomiques, notamment la stabilité macroéconomique, avec la mise en œuvre d'un filet de sécurité minimum.

*Une bonne réforme doit être transversale, cohérente avec les réformes précédentes et être menée sur la durée*, moyennant certaines inflexions suggérées par son évaluation. Pour qu'elle soit politiquement acceptable, elle doit être accompagnée du point de vue conjoncturel par l'État. Enfin, la réforme doit être adaptée à la situation macroéconomique et au contexte institutionnel à faire évoluer. Les évidences empiriques révèlent que les réformes réussissent là où les institutions sont performantes. Or, en RDC, fondamentalement, les institutions économiques et politiques sont à reconstruire[7]. Dans ce contexte, la mise en œuvre des réformes structurelles aura davantage besoin du caractère déterminé et visionnaire du réformateur qui doit privilégier les horizons de long terme, et résister aux différents groupes de pression qui œuvrent soit pour le statu quo, soit pour la préservation de leurs intérêts privés au détriment de ceux de la majorité.

Dans la section qui suit, nous proposons un cadre d'analyse susceptible de formaliser les interactions partisanes dans le processus de mise en œuvre des réformes structurelles.

## 3. L'offre et la demande du leadership

Le modèle que nous dérivons est une tentative dont la modeste ambition est d'essayer de formaliser les interactions partisanes entre un décideur politique et ses suiveurs. Considérons un cadre d'analyse où les réformes structurelles peuvent profiter soit à une majorité de la population participante (M), soit à une minorité ($\mu$). La probabilité de réussite d'une réforme structurelle ($\Psi$) dépend d'une part de la fraction de la population qui participe effectivement à la mise en œuvre de la réforme en cause et du degré de complémentarité entre participants et, d'autre part, du degré d'incertitude associé au processus de cette réforme. D'où:

$$\Psi = \frac{1}{\phi} \times a \times x^{\phi},$$ (1)

où $1 > a > 0$ dénote le degré d'incertitude lié au processus d'exécution de la réforme, $x$ la fraction de la population participante dans le processus de réforme, et $\phi > 1$ le degré de complémentarité entre participants. La figure 2 illustre comment les différents paramètres affectent la variation de de la fonction $\Psi$, toutes choses étant égales par ailleurs.

---

[7] Selon le rapport 2020 de *The Fund of Peace*, la RDC occupe le 5ème rang dans la liste des États faillis, après le Yémen, la Somalie, le Soudan du Sud et la Syrie respectivement.

Figure 2: Effet des différents paramètres sur Ψ, *ceteris paribus*

Panel (a): Effet de *a* sur \Psy, *ceteris paribus*

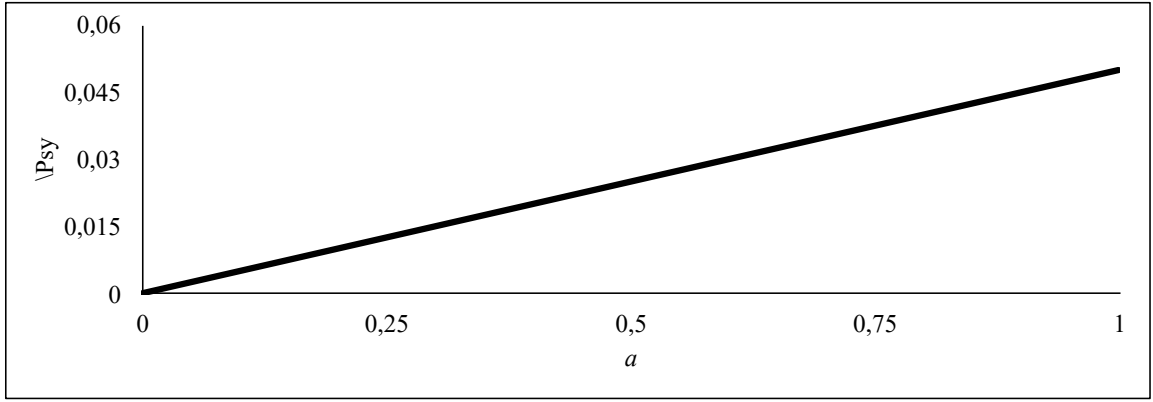

Panel (b): Effet de \phi sur \Psy, *ceteris paribus*

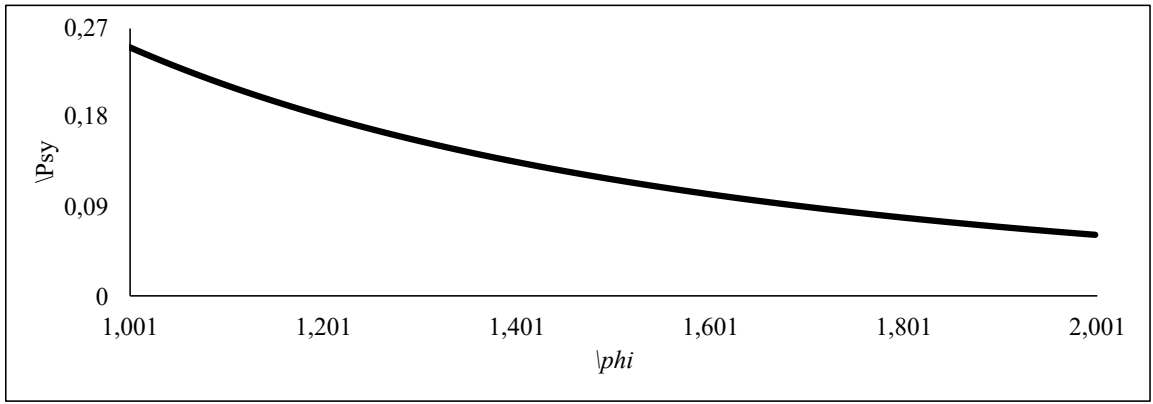

Panel (c): Effet de *x* sur \Psy, *ceteris paribus*

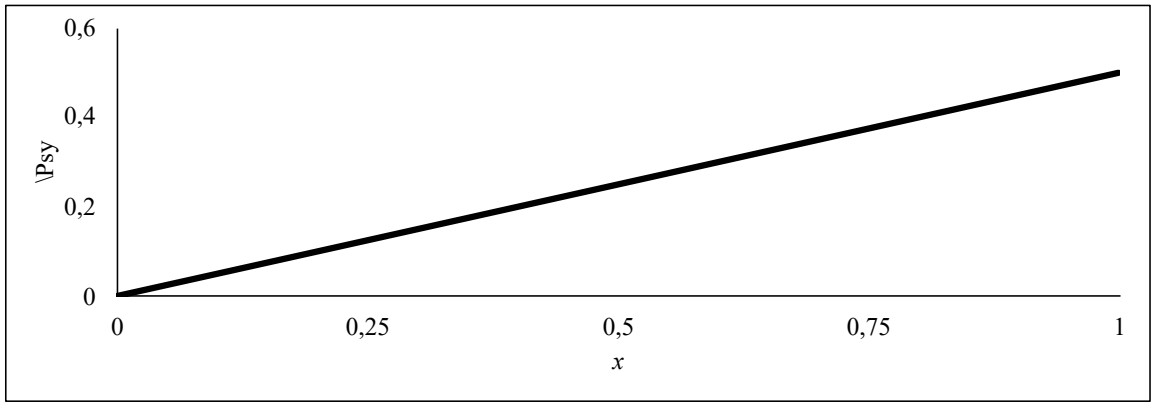

Note: Le degré d'incertitude $a$ varie entre 0 (incertitude totale) et 1 (certitude totale). Le degré de complémentarité $\phi$ varie entre 1 (forte complémentarité) et infini (faible complémentarité).

Sans perte de généralité, supposons qu'il existe trois états du monde: $E^{(1)}$, $E^{(2)}$ et $E^{(3)}$, respectivement avec les probabilités de survenance suivante: $p_1$, $(1-p_1)p_2$ et $(1 - p_1)(1 - p_2)$. L'incertitude dans le processus de réforme réside en ce qu'au niveau individuel, les participants ne disposent pas d'informations sur l'état sous-jacent du monde. Dans la fraction $x$ (cf. *Équation 1*), une sous-fraction des participants appartient au groupe majoritaire et l'autre sous-fraction complémentaire appartient au groupe minoritaire. Lors de la mise en œuvre d'une réforme, les participants de différents groupes (c'est-à-dire majoritaire et minoritaire) supportent le coût de sa mise en œuvre et, de par leur participation et implication, augmentent les chances de son succès. Cependant, les gains escomptés à l'issue d'une réforme dépendent de la réalisation des états du monde.

Tableau 1: Les états du monde

| État du monde | | Description | À qui reviennent les gains de la réforme? |
|---|---|---|---|
| $E^{(1)}$ | : | Le changement n'est pas possible (statu quo) | Uniquement aux groupes de pression (lobbyistes). |
| $E^{(2)}$ | : | Le changement est possible | À la minorité de la population, y compris le dirigeant politique à l'origine de la réforme[8]. |
| $E^{(3)}$ | : | Le changement est possible | À la majorité de la population, y compris la minorité et le dirigeant politique à l'origine de la réforme. |

Dans le premier état du monde, le changement est impossible. Dans ce cas, toute action ou tentative entreprise par un dirigeant politique allant dans le sens d'initier un ou des changements institutionnels aggraverait la situation de tous les joueurs, y compris celle du réformateur (i.e. du dirigeant politique à l'origine de la réforme). En revanche, dans les deuxième et troisième états du monde, le changement est possible. La différence entre les états $E^{(2)}$ et $E^{(3)}$ réside en ce que dans l'état $E^{(2)}$, les gains du changement profitent uniquement à la minorité, alors que dans l'état $E^{(3)}$, les gains d'un changement profitent à toute personne qui a participé activement au processus de la réforme, quel que soit son groupe d'appartenance (majorité ou minorité). Pour une raison ou une autre, les intérêts d'un dirigeant politique peuvent, en effet, être alignés sur l'un ou l'autre groupe d'intérêt. Lorsque les préférences d'un dirigeant politique sont de type « partisan », ce dernier ne s'opposera pas à ce que les changements institutionnels surviennent même dans l'état $E^{(2)}$, où seule la minorité de la population participante, y compris bien-sûr lui-même, profite des gains de la réforme en œuvre, quand bien même tous (majorité et minorité) en ont activement supporté le coût de réalisation[9]. En revanche, lorsque les préférences d'un dirigeant politique sont de type « non-partisan », ce dernier s'assure toujours, avant la mise en œuvre de toute réforme structurelle, que les gains du changement profiteraient à la majorité la population et, en conséquence, œuvrera acharnement pour que le changement se produise uniquement dans l'état $E^{(3)}$.

---

[8] L'état du monde $E^{(2)}$ peut aussi être vu comme un état où la réforme est personnellement bénéfique pour une classe politique, sans apporter de bénéfices généralisés. C'est le cas notamment lorsqu'une classe politique dirigeante décide de renforcer égoïstement son emprise sur le pouvoir.

La connaissance a priori de l'état de la nature aurait donc, sans doute, une influence sur le taux d'adhésion et de participation de la population dans le processus d'une réforme, et de ce fait sur la probabilité de sa réussite. Cependant, comme indiqué précédemment, la population n'a pas accès aux informations sur l'état sous-jacent du monde. C'est dans ce contexte que le rôle d'un « leader » devient important. En effet, dans notre modèle, nous supposons que le rôle du dirigeant politique, s'il est un leader, serait d'œuvrer dans l'acquisition de ces informations, puis de les transmettre à la population une fois effectivement acquises. Pour accéder aux informations sur l'état sous-jacent du monde, un dirigeant politique devra supporter le coût suivant:

$$c_\pi = \frac{1}{2} \times q \times \pi^2, \tag{2}$$

où $\pi$ dénote la probabilité qu'un dirigeant politique acquière des informations sur un état donné du monde et $q$ son habileté à identifier les opportunités de changement.

Soit $G_i$ le gain que la réussite du changement à l'état $E^{(i)}$, $i \in \{2,3\}$, procure au leader. Ce gain peut être matériel ou immatériel, et est tel que:

$$G_i < \frac{q}{(1-p_1) \times a \times \gamma}, \tag{3}$$

où $1 > \gamma > 0$ dénote la fraction de la population qui accède aux informations diffusées par un dirigeant politique. Pour un dirigeant politique partisan: $G_2 > 0$ et $G_3 > 0$; alors que pour un dirigeant politique non-partisan, c'est-à-dire un leader: $G_2 = 0$ et $G_3 > 0$.

Le coût pour un individu de participer à la mise en œuvre d'une réforme est soit nul (ou très fiable) avec une probabilité $\theta$, soit à valeur dans un support $[0, \kappa_{max}]$ distribué uniformément avec une probabilité $1 - \theta$. Le coût de participation ($\kappa_\theta$) pour les partisans (adeptes, militants ou « suiveurs ») est faible. En effet, ayant notamment adopté psychologiquement l'idéologie de leur chef de file ou « autorité morale », les partisans répondent activement à chaque invitation à participer dans les mouvements de masse ou, dans le cas d'espèce, dans un processus des changements ou des réformes, chaque fois cela leur est demandé. Dans ce cas précis, les partisans peuvent être considérés comme une fraction de la population qui est dévouée et fidèle à un dirigeant politique, lui offrant un soutien inconditionnel, et donc jouant un rôle pivot pour stimuler la mise en œuvre effective des réformes initiées leur chef de file. En revanche, pour les autres participants (« non-suiveurs »), $\kappa$ est distribué uniformément entre 0 et $\kappa_{max}$. Pour ces derniers, la décision de répondre activement à l'appel à l'action d'un dirigeant politique n'est pas automatique et dépend de plusieurs facteurs, notamment: (i) la crédibilité du dirigeant politique à l'initiative de la réforme, (ii) l'habileté du dirigeant à les persuader du bien-fondé collectif de la réforme, (iii) l'arbitrage entre leurs attentes quant à la réussite effective de la réforme envisagée et des coûts implicites à supporter.

---

[9] Ainsi, dans les pays en développement, si les changements institutionnels ne profitent qu'à la minorité de la population, c'est parce que fondamentalement les dirigeants politiques sont pour la plupart de cas « partisans », c'est-à-dire ils ne sont pas véritablement des « leaders ».

Nous supposons que la valeur maximale du coût à supporter, $\kappa_{max}$, est suffisamment élevée de sorte que toute la population ne pourra pas participer activement au processus du changement. D'où:

$$\Gamma < \frac{\kappa_{max}}{a \times \gamma}, \tag{4}$$

où $\Gamma > 0$ dénote le gain d'une réforme réussie pour un participant. Le coût de participation est tel que:

$$\kappa_\theta = \frac{1}{2} \times w \times \theta^2, \tag{5}$$

où $w$ dénote le coût ex post de participation. Plus un individu s'engage tôt, plus le gain espéré sera plus élevé.

## 4. Quelques implications du modèle à l'équilibre

Dans cette section, nous analysons les implications du modèle selon que: (i) les préférences du chef de file ou de l'autorité morale sont de type « partisan » (i.e. le dirigeant politique n'est pas un leader); les préférences du chef de file sont de type « non-partisan » (i.e. le dirigeant politique est un « leader »)[10].

Nous supposons que les préférences du chef de file sont de notoriété publique. Cette hypothèse élimine de ce fait l'existence des menaces non crédibles. L'analyse des implications du modèle à l'équilibre considère le timing suivant. À la période initiale (*t=0*): d'abord, chaque membre de la population estime combien cela lui coûterait de s'engager à un processus de changement. Ensuite, l'état du monde est réalisé. À la période *t=1:* le leader œuvre pour acquérir les informations sur l'état sous-jacent du monde. Une fois que quelques informations sur l'état sous-jacent du monde sont acquises, il les communique à la population.

---

[10] Pour rappel, lorsque les préférences d'un dirigeant politique sont de type « partisan », ce dernier ne s'opposera pas à ce que les changements institutionnels surviennent même dans l'état où seule la minorité de la population participante profite des gains de la réforme en œuvre. En revanche, lorsque les préférences d'un dirigeant politique sont de type « non-partisan », ce dernier s'assure toujours, avant la mise en œuvre de toute réforme structurelle, que les gains du changement profiteraient à la majorité de la population.

Finalement, à la période *t=2:* chaque membre de la population décide de participer ou non au processus de changement. Selon l'état du monde et la fraction de la population participante, le changement se produit ou non et les gains sont établis.

Tableau 2: Probabilités de survenance des états du monde

| État du monde | Probabilités | À qui reviennent les gains de la réforme? |
|---|---|---|
| $E^{(1)}$ | : $p_1$ | Uniquement aux groupes de pression (lobbyistes). |
| $E^{(2)}$ | : $(1-p_1)p_2$ | À la minorité de la population, y compris le dirigeant politique à l'origine de la réforme. |
| $E^{(3)}$ | : $(1-p_1)(1-p_2)$ | À la majorité de la population, y compris la minorité et le dirigeant politique à l'origine de la réforme. |

Considérons, tout d'abord, le cas d'un chef de file partisan, dont les préférences peuvent être distinctes de celles de la majorité. Résolvons le problème en appliquant l'induction à rebours. À la période *t=2*, un membre de la population qui n'a pas reçu les informations sur l'état de la nature communiquées par le leader ne participera pas au processus de changement quel que soit le coût que cela implique. En revanche, la fraction des membres qui ont reçu le message peuvent décider de participer ou non.

Soit $1 > s > 0$ la probabilité que les préférences du chef de file coïncident avec celles de la majorité. Il s'agit, en effet, une probabilité qui découle de la perception qu'a la population sur les préférences du chef de file. Dès lors, après que la fraction $\gamma$ de la population ait reçu l'appel $G$ à s'engager pour le changement, la perception qu'ont les personnes qui appartiennent au groupe M (i.e. la majorité) est donnée par la probabilité conditionnelle suivante:

$$\mathbb{P}\big(E^{(3)}/G\big) = \frac{p_2}{p_2 + (1-s) \times (1-p_2)}, \tag{6}$$

où $\mathbb{P}\big(E^{(3)}/G\big)$ dénote la probabilité que l'état sous-jacent du monde est $E^{(3)}$ étant donné le message $G$ émis par le chef de file. Dans la fraction $\gamma$ de la population qui a reçu, une sous-fraction $\theta$ est constituée des partisans engagés. Ces derniers, n'ayant qu'un coût de participation très faible ($\kappa = 0$), vont sûrement répondre activement à l'invitation du leader pour le changement.

Pour la fraction complémentaire $(1 - \theta)$, seules les personnes dont le coût est inférieur ou égal à $\kappa^*$, $\kappa^* \in [0, \kappa_{msx}]$, vont participer:

$$\kappa^* = \theta \times \left[ \frac{1}{a\gamma} \frac{1}{\Gamma} \frac{1}{\mathbb{P}(E^{(3)}/G)} + \frac{1-\theta}{\kappa_{max}} \right]^{-1}. \tag{8}$$

Cette équation détermine la « demande du leadership » pour les non-suiveurs au cas où le chef de file n'est pas un leader.

Figure 3: Offre et demande de leadership

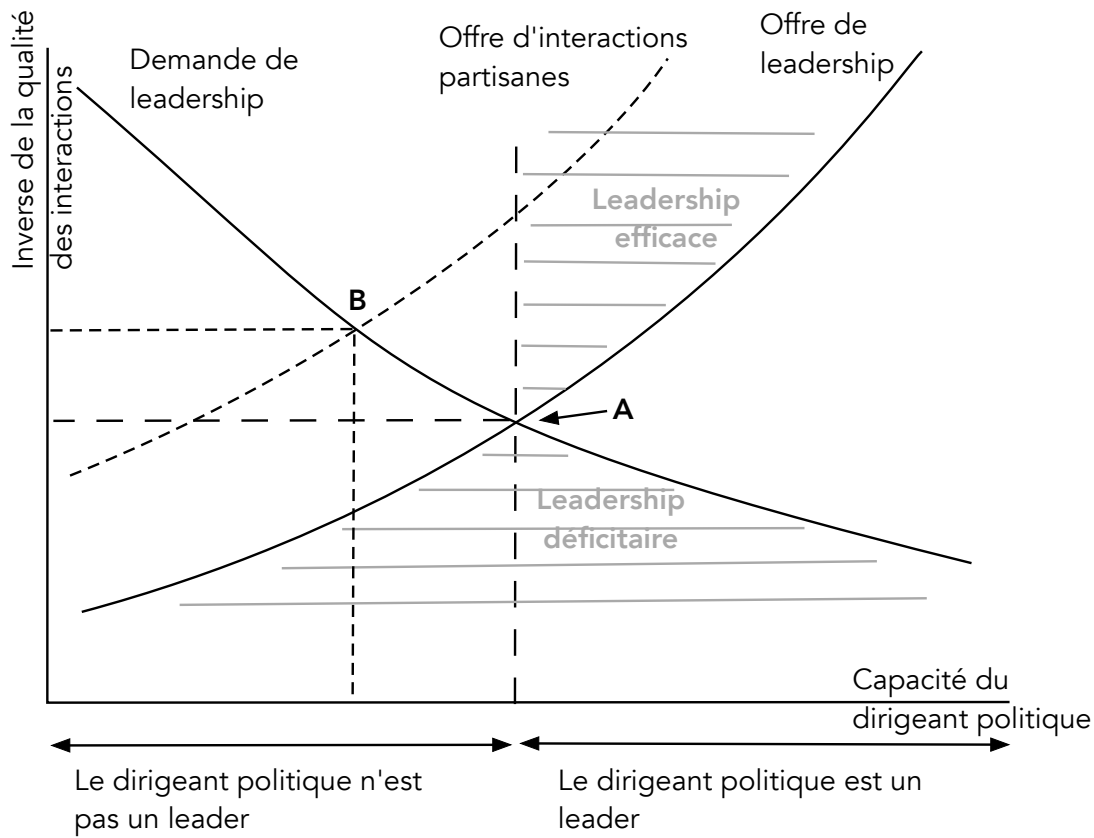

Note: Point **A**: Équilibre entre l'offre de leadership et la demande de leadership. Point **B**: Équilibre entre l'offre d'interactions partisanes et la demande de leadership.

Considérons à présent que le chef de file est un leader. Une fois de plus, résolvons le problème en appliquant l'induction à rebours. À la période *t=2*, un membre de la population qui n'a pas reçu les informations sur l'état de la nature communiquées par le leader ne participera pas au processus de changement quel que soit le coût que cela implique. En revanche, la fraction des membres qui ont reçu le message peuvent décider de participer ou non. Pour un suiveur, le coût étant faible ($\kappa = 0$), il participera toujours.

En revanche, pour un non-suiveur, $\kappa \in [0, \kappa_{max}]$, il ne s'engagera qu'après avoir pesé les coûts et les gains de sa participation. Il choisira de participer uniquement si :

$$\kappa \leq a \times \Gamma \times \mathbb{E}\{x\}, \tag{9}$$

où le terme $\mathbb{E}\{x\}$ dénote l'attente du non-suiveur quant au nombre de personnes qui participeront au processus de changement. Autrement dit, il s'agit de l'espérance mathématique de la fraction de la population participante dans le processus de réforme À l'équilibre :

$$\kappa^* = a \times \Gamma \times \mathbb{E}\{x\}. \tag{10}$$

Puisque dans ce cas, le dirigeant politique est non-partisan, les participants déduiront que l'état sous-jacent du monde ne peut qu'être $E^{(3)}$. En effet, contrairement à un dirigeant politique partisan, en aucun cas, un leader ne peut initier une réforme lorsque l'état sous-jacent du monde est $E^{(1)}$ ou $E^{(2)}$.

Étant donné la capacité du leader à communiquer, une fraction $\gamma$ de la population le reçoit. De ce nombre, une sous-fraction $\theta$ est constituée des partisans engagés. Ces derniers, n'ayant qu'un coût de participation très faible ($\kappa = 0$), vont sûrement répondre activement à l'invitation du leader pour le changement. Pour la fraction complémentaire $(1 - \theta)$, seules les personnes dont le coût est inférieur ou égal à $\kappa^*$ vont participer. Dès lors, en agrégeant, la fraction totale personnes qui vont effectivement participer au processus de changement est telle que :

$$x = \gamma \times \left[\theta + (1 - \theta)\frac{\kappa^*}{\kappa_{max}}\right]. \tag{11}$$

Enfin, étant donné la réalisation des attentes individuelles concernant la participation globale, la décision de participation à l'équilibre devient :

$$\kappa^* = \theta \times \left[\frac{1}{a\gamma}\frac{1}{\Gamma} + \frac{1 - \theta}{\kappa_{max}}\right]^{-1}. \tag{12}$$

L'équation (12) détermine la « demande du leadership » pour les non-suiveurs au cas où le chef de file est un leader. Ceteris paribus (notamment $a$), la décision ou le degré de participation pour les non-suiveurs est croissance en $\gamma$ (l'efficacité du leader à communiquer), en $\theta$ (le nombre des partisans)[11], et décroissante en $\kappa_{max}$ (le coût maximal de participation). En outre, le gain attendu à l'issue d'une réforme est positivement corrélé à la décision de participation pour les non-suiveurs. In fine, étant donné que la probabilité de réussite d'une réforme est fonction de la fraction de la population participante dans le processus de réforme, les variations des paramètres qui augmentent $\kappa^*$, notamment $\{\theta, a, \gamma, \Gamma\}$, augmentent conséquemment la probabilité de réussite d'une réforme.

---

[11] Il est intéressant de noter, ici, qu'à mesure que le nombre d'adeptes engagés augmente, la participation du reste de la population augmente aussi. Ainsi, le fait d'avoir un noyau de partisans ou adeptes engagés est nécessaire pour avoir une chance plus élevée de réussir une réforme.

À la lumière du cadre d'analyse que nous venons de développer et sur base de la documentation existante, nous nous proposons, dans la section qui suit, d'illustrer à l'aide de quelques exemples typiques les chances de réussite des réformes institutionnelles lorsque les décideurs politiques font preuve d'engagement et de leadership dans leur mise en œuvre. Nous considérons le cas de la RDC.

## 5. Leadership et réformes institutionnelles en RDC: un diagnostic détaillé

**Background**. Dans ce papier, nous partons de l'idée que l'efficacité d'un dirigeant politique dépend essentiellement des quatre facteurs-clés: (i) son *habileté* à identifier, dans des circonstances et moments appropriés, les opportunités de changement; (ii) sa capacité à attirer et à conserver un groupe des collaborateurs et des partisans engagés; (iii) son *habileté* à communiquer efficacement avec la population; (iv) la perception qu'a le public de la capacité d'un dirigeant politique à initier et à mener avec succès les réformes. Le modèle développé a montré comment ces facteurs affectent le processus de mise en œuvre des réformes structurelles, notamment la probabilité de réussite d'une réforme structurelle.

Dans la présente section, nous considérons les évidences tirées de l'histoire économique récente de la RDC pour montrer deux faits marquants: (i) le leadership institutionnel a pris corps dans l'impulsion et la coordination des réformes structurelles de grande envergure à partir des années 2001, à la suite notamment de la reprise de la coopération avec les institutions de Bretton Woods, notamment le Fonds monétaire international (FMI) et la Banque mondiale; (ii) ce leadership dans la mise en œuvre de réformes structurelles ont connu un point d'inflexion et un nouvel élan en 2010[12], avant de connaître une décélération à partir de 2016[13].

---

[12] Avec notamment le renforcement du ministère des finances en tant que seul centre d'ordonnancement tant des recettes que des dépenses de l'État et la gestion des opérations financières de l'État sur la base de deux instruments principaux, à savoir, le plan de trésorerie et le plan d'engagement budgétaire.

[13] Une décélération expliquée notamment par la succession des gouvernements de cohabitation et de coalition. La succession des gouvernements de cohabitation et de coalition en RDC, durant la période 2016-2019, s'explique principalement par le report des élections présidentielles, législatives et provinciales initialement prévues en 2016 et des tensions qui s'en sont ensuivies. Le Gouvernement Matata, en place depuis 2012, a été successivement relayé par: (i) le Gouvernement Badibanga (19 décembre 2016-9 mai 2017) issu de l'accord politique de la cité de l'OUA; (ii) le Gouvernement Tshibala (9 mai 2017-6 septembre 2019) issu de l'accord politique global et inclusif (APGI), appelé également Accord de la Saint-Sylvestre. Les deux accords politiques, ont été respectivement signés le 18 octobre 2016 au siège de la cité de l'OUA sous la médiation de l'ex-premier ministre togolais Edem Kodjo, désigné « facilitateur » du dialogue par l'Union africaine (UA), et le 31décembre 2016 au Centre Interdiocésain de Kinshasa sous la médiation de la Conférence Épiscopale Nationale de la République démocratique du Congo (CENCO). Ces accords ont consacré le dialogue et la cohabitation entre la majorité présidentielle, et une frange de l'opposition et de la société civile jusqu'à la tenue des élections présidentielles, législatives et provinciales fixées en décembre 2018. Le Gouvernement Ilunga (6 septembre 2019-à ce jour) est

Ainsi, pour comprendre les interactions entre le leadership et la conduite des réformes en RDC, nous nous focaliserons sur la période 2010-2016[14].

Dans l'objectif de soutenir la croissance économique dans un environnement avec une marge des fluctuations d'inflation maîtrisée; dès 2010, le ministère des finances avait institué comme modalité permanente: la gestion sur la base caisse. Cette décision a rigoureusement conditionné l'exécution des dépenses à l'existence préalable des recettes effectivement recouvrées ou des marges de trésorerie préalablement constituées au niveau du compte général du Trésor, logé à la Banque centrale du Congo (BCC). Par ailleurs, des instructions strictes étaient données pour le respect des étapes budgétaires de la chaîne de la dépense allant de l'engagement, en passant par la liquidation, l'ordonnancement et le paiement. Les dépenses hors chaine étaient réduites au minimum et ne pouvaient concerner que des dépenses urgentes relevant de la souveraineté. Ces dernières étaient logées dans le compte *ad hoc* ouvert dans les livres de la BCC et devaient être régularisées dans les quarante-huit heures par l'envoi, par le ministère des finances d'un ordre de paiement informatisé *ad hoc*.

Les dépenses d'investissement devraient veiller à l'adéquation entre l'exécution financière et physique des projets dans le cadre des marchés publics *ad hoc*. En ce qui concerne les transferts aux provinces au titre de rétrocession, ils ne pouvaient être libérés qu'après présentation des justificatifs *ad hoc* d'utilisation conforme des transferts antérieurs. Quant à la dette intérieure, les remboursements étaient conditionnés par leur certification préalable, l'application, après négociation entre le gouvernement et les créanciers, de la décote à appliquer conformément aux clauses afférant au traitement égalitaire (*pari passu*).

Au sujet de la paie des agents et fonctionnaires de l'État, il a été décidé de la suppression de la paie par les comptables d'État qui retiraient les fonds dans les banques et procédaient à sa liquidation selon leur bon vouloir. La paie était désormais effectuée par voie bancaire et libérée avant le 20 de chaque mois (*la bancarisation de la paie*).

S'agissant de la mobilisation des recettes, des réunions régulières de 60 minutes étaient tenues avec les régies, chaque mardi à 7 heures, pour évaluer leurs performances par rapport aux assignations leur conférées. Les rétrocessions étaient faites au *prorata* des

---

issu de la coalition entre les forces politiques du candidat vainqueur de l'élection présidentielle du 30 décembre 2018 et celles de son prédécesseur qui ont préservé la majorité au parlement et dans les 26 provinces du pays (gouvernorat). Aussi, à souligner que, plus tôt, en date du 7 décembre 2014, le Gouvernement Matata avait connu un réaménagement technique à l'effet de mettre en exécution la principale conclusion du Comité de suivi des Concertations nationales, co-présidé par Aubin Minaku et Léon Kengo wa Dondo (respectivement présidents de la chambre basse et de la chambre haute du parlement en RDC), qui recommandait la formation d'un gouvernement de cohésion nationale dans l'objectif de planifier le développement socio-économique dans la paix et la concorde.

[14] Réformes impulsées et conduites par le professeur Matata Ponyo Mapon, alors respectivement ministre des finances (2010-2012) et premier ministre (2012-2016) de la RDC.

recettes effectivement mobilisées. Dans ce cadre, plusieurs taxes redondantes ont été supprimées. Les échéances fiscales ont été précisées.

Ce nouveau leadership dans la conduite de la politique budgétaire a permis non seulement l'amélioration de la qualité de la dépense et la mobilisation efficace des recettes, elle a aussi déterminé le respect régulier des critères quantitatifs pertinents des avoirs extérieurs nets, des avoirs intérieurs nets et du crédit net à l'État et la restauration d'une véritable stabilité du cadre macroéconomique, favorisant rapidement l'atteinte du point d'achèvement de l'Initiative PPTE en juin 2010 et l'avènement d'une croissance forte entre 2010 et 2015.

**Les séminaires gouvernementaux et les lettres de mission**. Dans le cadre de la mise en place d'institutions efficaces, le premier ministre a pris l'initiative d'organiser un séminaire gouvernemental du 03 au 04 juillet 2012. Une première expérience du genre dans l'histoire gouvernementale congolaise. La réunion de l'ensemble du gouvernement pendant deux jours a représenté un investissement en temps considérable pour l'ensemble des ministres, marqué par la richesse des débats entre plus de 40 participants. Le but principal de ce séminaire était de favoriser la prise de conscience par l'ensemble du gouvernement de l'ampleur des défis auxquels est confronté le pays et de la nécessité d'une approche collective face à ces défis, pour tenter d'y répondre en sortant d'une logique qui serait fondée sur une simple juxtaposition d'approches sectorielles peu ou non coordonnées. Pour réexaminer la récurrence de certaines qui plombent le développement économique des pays en développement comme la RDC et tenter de les aborder sous des angles différents, ce séminaire a connu la participation d'experts étrangers, notamment les chercheurs du Korean Development Institute (KDI), le premier ministre honoraire Tertius Zongo de la république du Burkina Faso, et un ancien Directeur de la Banque mondiale, Serge Michaelof. En sus, un second séminaire du gouvernement fut organisé à Kinshasa à la Cité de l'Union africaine du 12 au 13 février 2015 à la suite d'un remaniement ministériel. Ces séminaires ont permis à chaque fois d'assurer un même niveau d'information à tous les membres de l'Exécutif et ainsi faciliter la prise de décisions consensuelles dont le suivi et l'évaluation ont été faits à l'aide des lettres de mission.

Au-delà du recours à l'instrument budgétaire, et pour en assurer une réelle efficacité opérationnelle, il était urgent pour le gouvernement de renforcer la coordination de son action en vue d'en garantir des résultats concrets. Ainsi, le *nouveau consensus sur les institutions efficaces* pour le développement a amené le gouvernement à considérer les *réformes institutionnelles dans le secteur public* en vue de libérer les externalités positives propices au meilleur suivi de l'action gouvernementale. Se conformant à la notion développée de PIE (plate-forme des institutions efficaces) par la Banque mondiale dans le cadre de l'amélioration de l'efficacité institutionnelle, le gouvernement a, à partir de 2012, pris des initiatives visant le renforcement des capacités de suivi des ministères, avec une meilleure coordination au cabinet du premier ministre. Ce qui a peu à peu permis de migrer vers des institutions publiques responsables, inclusives et transparentes capables de mettre en œuvre des politiques publiques et une gestion efficace des ressources pour des services publics durables.

Par ailleurs, durant la même période, des avancées ont été notées dans les domaines suivants: l'ouverture à la concurrence de plusieurs secteurs d'activités jadis monopolistiques par le dépôt pour adoption de textes de lois dont la loi sur l'agriculture, la loi sur l'électricité, le code des assurances, la loi relative au crédit-bail, la loi sur le statut des fonctionnaires avec possibilité de recourir à des experts externes pour des postes précis, le code des hydrocarbures, la loi sur l'eau, la loi relative au partenariat public-privé.

En conséquence, en 2013, lors de l'évaluation des politiques et institutions nationales, par le *Country Policy Institutional Assessment* (CPIA) de la Banque mondiale, la notation de la RDC a progressé de 2,7 à 2,9. Ce qui a fait dire à cette organisation internationale qu'une telle progression a été unique par son ampleur. Car la RDC a été le seul des 82 pays de l'Association Internationale pour le Développement (IDA) à avoir affiché une telle progression en 2013. Dans le même temps, l'amélioration du climat des affaires a placé la RDC parmi le top 10 des pays réformateurs au classement *Doing Business* de la Banque mondiale en 2013.

Faute de culture d'obligation de rendre compte (redevabilité), les ministres n'étaient auparavant jamais tenus pour responsables de leurs actes. D'où, l'introduction des lettres de mission qui leur assignaient des cibles opérationnelles précises; lesquelles étaient périodiquement évaluées. Ces lettres de mission ont notamment joué trois rôles majeurs, en tant que: (i) outil d'amélioration de gestion budgétaire; (ii) instrument d'évaluation des capacités institutionnelles; (iii) dispositif de redevabilité et d'inclusion.

La modernisation de la gestion des finances publiques a amené les pouvoirs publics à s'engager dans la budgétisation fondée sur une logique de résultats. La loi no. 11/011 du 13 juillet 2011 relative aux finances publiques a introduit une dimension nouvelle dans la gestion des finances publiques par le passage d'une gestion centrée sur les moyens à une gestion axée sur les résultats (GAR). La nouvelle loi des finances publiques impose donc à tout gestionnaire des crédits une obligation de résultats, pour l'atteinte des objectifs de développement. Désormais, l'action publique doit répondre à une gestion axée sur les résultats, et les ressources budgétaires sont allouées par axe stratégique, au profit d'actions à mener dans le cadre des politiques publiques, en privilégiant la qualité de la dépense publique. Suivant les dispositions des articles 20, 43 et 82 de la loi relative aux finances publiques, toutes les ressources du budget traduisent le plan d'actions du gouvernement dont les objectifs visés et les résultats attendus font l'objet d'une évaluation au moyen d'indicateurs de performance.

Ce choix a été intégré par le gouvernement dans son programme d'action 2012-2016 (PAG). Le gouvernement s'est assigné plusieurs objectifs alignés aux piliers du DSCRP. Dans le chapitre 6 dudit PAG consacré au « dispositif de suivi et évaluation », le gouvernement a levé l'option de mettre en place un système efficace de suivi et évaluation pour évaluer, à des intervalles de temps régulier, le chemin parcouru et prendre si nécessaire les mesures correctives. La gestion axée sur les résultats est désormais privilégiée pour mesurer les performances de chaque secteur. Les ressources sont allouées par axe stratégique en privilégiant la qualité de la dépense publique. La préoccupation majeure a été de migrer vers une centralisation et une coordination étroite de la planification stratégique du développement intégrée au processus budgétaire. Le gouvernement a ainsi posé les jalons pour la mise en place d'un système performant de suivi évaluation de l'action gouvernementale grâce aux lettres de mission. Dès l'approbation du programme du gouvernement par l'Assemblée nationale, une feuille de route des 100 premiers jours du gouvernement a été rendue publique en date 22 mai 2012. Sur la base des feuilles de route pour la mise en œuvre du programme d'action du gouvernement 2012-2016, une évaluation de l'action des ministères pour les 100 premiers jours a été publiée en date du 17 août 2012.

Tirant expérience de cet exercice, et en application des recommandations du séminaire gouvernemental**,** des lettres de mission ont été préparées pour le dernier trimestre de l'année 2012, à l'attention de tous les membres du gouvernement et remises au cours d'une cérémonie solennelle le 28 septembre 2012. Il s'agit d'un pas important vers une meilleure coordination de l'action gouvernementale et de l'évaluation du rendement des administrations. Ainsi, sur le plan du mode opératoire interne, l'élaboration des « lettres de mission » fixant les priorités sectorielles annuelles pour chaque membre du gouvernement a permis non seulement un renforcement de la cohérence de l'action gouvernementale, une consolidation de la coordination gouvernementale, mais aussi une gestion véritablement axée sur les résultats.

De ce fait, conformément à la lettre circulaire du premier ministre no. CAB/PM/CR/JPM/ 2014/16612 du 15 décembre 2014 qui organise le processus d'élaboration des lettres de mission, les priorités formulées par les ministères sectoriels doivent être communiquées par les ministres gestionnaires des crédits aux services compétents du ministère du budget pour leur inscription aux plans d'engagement budgétaire (PEB) trimestriels, en veillant aux équilibres du plan de trésorerie (PTR), en exécution de la circulaire no. 001/ME/MIN.BUDGET/2015 du 05 janvier 2015 contenant les instructions relatives à l'exécution de la loi de finances no. 14/027 du 31 décembre 2014. Lors du suivi, il s'agit de vérifier l'atteinte des résultats (GAR) par l'exécution du budget conformément aux PEB harmonisés avec le PTR annuel. Dès lors, les lettres de mission permettent d'évaluer le progrès accomplis et les capacités institutionnelles. Après la phase pilote de 2012, un mode de suivi et évaluation stricte a été mis en place par la lettre circulaire du premier ministre no. CAB/PM/CR/JPM/ 2013/2699 du 08 mai 2013 au regard des principaux objectifs du Programme du gouvernement à savoir: (i) poursuivre et finaliser les réformes institutionnelles en vue de renforcer l'efficacité de l'État; (ii) consolider la stabilité du cadre macroéconomique et relancer la croissance et la création d'emploi; (iii) poursuivre la construction et la modernisation des infrastructures de base: routes, voiries, chemins de fer, voies d'eau, ports et aéroports, écoles et hôpitaux; (iv) améliorer les conditions de vie des populations congolaises; (v) renforcer le capital humain et faire de la société congolaise un vivier de la nouvelle citoyenneté; (vi) renforcer la diplomatie et la coopération au développement.

D'autres actions nouvelles étaient mises en œuvre pour améliorer la capacité opérationnelle de la primature et de l'ensemble de l'équipe ministérielle. Concernant la Primature, un module d'évaluation mensuelle des performances de l'ensemble du Cabinet du Premier ministre a été mis en place par le cabinet *PriceWaterhouse and Coopers.* S'agissant du gouvernement, tous les ministres étaient soumis à une cotation trimestrielle sur la base des performances dans la mise en œuvre des lettres de missions leur imparties. Au préalable, tous les ministres ont été invités à formuler les actions et mesures à suivre faisant ressortir clairement les politiques sectorielles en indiquant: (i) les priorités stratégiques par ministère pour l'année 2013; (ii) les résultats mesurables à atteindre au cours de l'année; (iii) les activités à réaliser aux fins d'atteindre lesdits résultats; (iv) les délais de réalisation pour chaque activité retenue.

Toutefois, dans le souci de prévenir toute évaluation subjective des performances, une méthodologie claire a été mise en place sous format matriciel faisant ressortir les activités retenues pondérées selon qu'elles: (i) concourent directement à la réalisation des priorités stratégiques sectorielles et des résultats attendus (« principale », 60%); (ii) contribuent à la préparation et à la planification des résultats (« secondaire », 30%); (iii) relèvent du fonctionnement normal du ministère sans impact immédiat dans l'atteinte des résultats (« tertiaire », 10%). Ces propositions ont fait l'objet d'une validation en Conseil des ministres, conformément aux dispositions des articles 27, 46 et 48 de l'Ordonnance no. 12/007 du 11 juin 2012 portant organisation et fonctionnement du gouvernement. Par ailleurs, chaque Membre du gouvernement a été invité à faire preuve de créativité et de volontarisme dans la préparation et la mise en œuvre desdites mesures. Pour la pérennisation de cet outil, des compilations annuelles des lettres de mission ont été transmises aux deux chambres du Parlement, à la Cour des comptes ainsi qu'à l'Inspection générale des finances en vue d'un suivi indépendant. Les lettres de mission ont permis la facilitation de la redevabilité et de l'inclusion tandis que le recentrage des institutions a favorisé la mise en place d'un cadre institutionnel nouveau.

Deux approches ont été utilisées par le gouvernement, au cours de la période 2012 à 2016, pour s'assurer de la validité, dans la transparence et l'inclusivité, des résultats du Programme économique, à savoir les missions indépendantes des experts internationaux et le suivi-citoyen. Soucieux de se soumettre à l'obligation de transparence, le gouvernement s'est astreint à rendre compte à travers un système indépendant efficace de suivi et évaluation les résultats du Programme d'action 2012-2016, approuvé par l'Assemblée nationale. Cette option visait à évaluer régulièrement le chemin parcouru et prendre, si nécessaire les mesures correctives. Dans le même ordre d'idées et loin de se complaire dans l'autosatisfaction, le gouvernement a accueilli des missions régulières d'experts internationaux qui ont séjourné à Kinshasa tous les trois mois d'octobre 2012 à janvier 2014. Les experts ont pu faire le suivi des recommandations opérationnelles des séminaires gouvernementaux. Ils ont ainsi pu accompagner la préparation d'une revue indépendante du Programme d'action du gouvernement pour évaluer les performances dans divers secteurs et proposer des mesures d'ajustement. Dans le cadre de l'implication des administrations pérennes dans le suivi de la performance des lettres de mission, le secrétariat général à la primature, à travers sa direction de suivi et évaluation des politiques (DSEP), a bénéficié du renforcement des capacités en GAR et pratiqué des exercices de suivi des lettres de mission en 2013.

Dès lors, il s'est avéré important d'élargir le champ d'intervention non seulement au niveau des administrations ministérielles sectorielles, mais également au niveau de la société civile pour une meilleure appropriation de la GAR et s'assurer de la participation citoyenne à travers des évaluations régulières de l'impact de l'action gouvernementale dans la mise en œuvre des lettres de mission 2015-2016. En effet, en appui aux activités de renforcement des capacités financées par la Banque mondiale à travers le Projet de Renforcement des Capacités en Gestion des Fonctions de base de l'Administration Publique (PRC-GAP) en faveur des Secrétariats généraux, il était désormais prévu l'implication de la société civile au suivi de l'action gouvernementale. La société civile sectorielle, encadrée par quatre ONG ayant des expériences dans le suivi des politiques et budgets publics, était appelée à participer aux efforts du gouvernement en s'appropriant les outils de suivi et évaluation des politiques publiques. Parmi ces ONG faîtières figurent le REGED (Réseau Gouvernance Économique et Démocratie) et l'OLCAC (Observatoire de Lutte contre la Corruption en Afrique Centrale). Ils étaient complémentés par deux autres ONG ayant une expérience similaire. Pour ce faire, les ONG ont commencé à prendre une part active aux ateliers de renforcement des capacités des secrétariats généraux en GAR et suivi des performances sectorielles des ministères. Cette démarche visait à garantir la participation citoyenne à la compréhension des processus d'élaboration des lettres de mission et leur finalité.

Très rapidement, le suivi régulier des priorités inscrites dans les lettres de mission a permis d'améliorer l'efficacité de l'action publique. Dans ce registre, on peut, en effet, citer plusieurs réformes institutionnelles qui stagnaient depuis plusieurs années: la réforme de l'OHADA. Pour une réforme lancée en 2003 dans le cadre de la reprise des programmes avec le FMI (PEG I), réforme qui a suscité beaucoup de réticences et de résistances dans le milieu des professionnels du droit. En 2012, le gouvernement a fini par se coaliser et obtenir l'appui du Parlement pour procéder au dépôt des instruments de ratification du Traité OHADA, soit près de dix ans après sa proposition. Il en est de même pour l'adhésion de la RDC à la Convention de New York de 1958 sur l'exécution des sentences arbitrales rendues à l'étranger.

En vue de consolider les acquis de l'ajustement budgétaire quantitatif, accélérer la croissance et lever les obstacles passés, le gouvernement a résolu de maintenir ce cap tout en redynamisant certaines réformes institutionnelles dans les principaux domaines de l'action publique. Deux d'entre elles méritent d'être examinées: (i) la bancarisation de la paie des agents et fonctionnaires de l'État; (ii) la modernisation de la gestion des ressources humaines de l'administration publique. Ces deux réformes apparaissent comme des innovations majeures ayant soutenu le nouveau cadre institutionnel au cours de la période allant de 2010 à 2016.

**La bancarisation**. En vue d'instaurer une véritable gouvernance dans la gestion des dépenses de rémunération qui absorbent près de 40% des recettes publiques mensuelles, le gouvernement a lancé la réforme de la bancarisation de la paie des fonctionnaires. Cette réforme, initiée de manière très volontariste par le gouvernement dès 2011, a bénéficié du soutien important des agents de l'État notamment grâce à l'*effet de dotation des agents économiques* mis en avant par Daniel Kahneman. Dans le cas d'espèce, cet effet a entraîné un réel renforcement des pouvoirs des fonctionnaires bénéficiaires immédiats de la réforme avec comme corollaire la fragilisation simultanée des rentiers, désormais privés de la manipulation manuelle de l'enveloppe salariale.

La bancarisation de la paie des agents et fonctionnaires de l'État est une réforme qui touche à la fois la gouvernance dans la gestion de l'administration publique ainsi que dans la gestion de finances publiques. En effet, cette réforme vise la maitrise des effectifs des agents et fonctionnaires de l'État avec en conséquence la maitrise de la masse salariale. Elle s'inscrit dans le cadre des réformes institutionnelles prônées par le gouvernement en vue de renforcer la gouvernance et l'efficacité de l'État. Si cette réforme institutionnelle a pour objectif fondamental une plus grande maîtrise des effectifs et de la masse salariale, il y a lieu d'expliquer qu'elle a visé à répondre de manière conjoncturelle aux dysfonctionnements chroniques dans la programmation budgétaire de la paie et assurer le respect du critère de gestion budgétaire relatif à la non- accumulation des arriérés de salaires. Sur le plan pratique, cette réforme consiste au paiement des salaires par voie de compte bancaire avec la collaboration des banques commerciales qui se sont constituées en véritables partenaires du gouvernement. Ces dernières se sont engagées au préalable à ouvrir des comptes individuels en faveur des fonctionnaires bénéficiaires de la réforme. Jusqu'alors, le paiement des rémunérations reposait sur des opérations en espèces entre les comptables publics et les salariés de l'État.

Sur le plan technique, la bancarisation de la paie implique l'harmonisation préalable du circuit de la paie des agents et fonctionnaires de l'État. Cette migration d'un système de manipulation matérielle de la monnaie fiduciaire par des comptables publics à des opérations comptables et bancaires a dû se faire de manière graduelle et progressive. La réforme de la bancarisation a été initiée en août 2011 avec la paie des institutions politiques pour un effectif de 2 444 unités. En juillet 2012, le gouvernement a accéléré le rythme de la réforme avec 141 785 unités. La bancarisation de la paie des agents et fonctionnaires de l'État en provinces est intervenue au mois d'octobre 2012, soit 15 mois après le lancement de la réforme à Kinshasa. Ce temps a été nécessaire à la consolidation des acquis de la réforme dans la capitale et à l'amélioration du suivi de sa mise en œuvre afin d'éviter de transplanter les insuffisances observées au départ dans les autres points prévus dans le chronogramme.

À terme, les effectifs à bancariser sont ceux de l'administration publique et des institutions correspondants aux plafonds des autorisations des emplois budgétaires rémunérés (actifs et passifs) de 1 143 908 unités inscrites à l'annexe 14 de la Loi de finances pour l'exercice 2016, pour une masse salariale de 2 045 173 009 335 de FC. Cette réforme permettra au gouvernement de maîtriser les effectifs et la masse salariale par la consolidation du fichier unique de tous les fonctionnaires de l'État. Ce fichier référentiel unique des agents de l'État constitue la première étape de la modernisation de l'administration publique et de l'élaboration d'une politique salariale cohérente et plus motivante pour les fonctionnaires et agents de l'État.

En réalité, la bancarisation de la paie des agents et fonctionnaires de l'État est le premier jalon dans la restauration de la gouvernance tant dans la gestion administrative des effectifs que dans la discipline budgétaire et l'orthodoxie financière en matière de dépenses salariales. En effet, étant donné que les dépenses de rémunérations absorbent près de 40% des ressources budgétaires internes, l'amélioration de la gouvernance dans l'exécution de dépenses de rémunérations amènerait à rationaliser près de la moitié de dépenses publiques, tel qu'il ressort de l'évolution des dépenses de rémunérations du document de Programmation budgétaire des actions du gouvernement 2015-2017.

Tableau 3: Tableau des effectifs bancarisés de 2011 à 2016

|  | 2011 | 2012 | 2013 | 2014 | 2015 | 2016 |
|---|---|---|---|---|---|---|
| Effectifs bancarisés | 26 871 | 295 601 | 612 069 | 669 066 | 683 812 | 747 093 |
| Effectifs totaux des agents actifs | 874 559 | 874 559 | 874 142 | 904 877 | 904 877 | 904 877 |
| Taux de bancarisation (%) | 3,1 | 33,8 | 70,0 | 73,9 | 75,6 | 82,6 |

Toutefois, des faiblesses structurelles et conjoncturelles continuent à peser sur l'aboutissement de cette réforme qui devrait produire son impact optimal à terme. La mesure structurelle la plus importante à finaliser porte sur la mise en réseau par fibre optique du système intégré de GRH-Paie déployé au ministère de la Fonction publique, gestionnaire des ressources humaines de l'État. La gouvernance et la transparence dans la gestion du fichier unique de la paie passe, en effet, par la mise en place de cette interconnexion entre les ministères de la Fonction publique, du budget, des finances et de la Banque centrale du Congo. Cette interconnexion devrait permettre de résoudre notamment certaines pesanteurs systémiques dont: (i) la lenteur dans le transfert des listings de la paie vers les banques commerciales; (ii) le problème lié aux mises à jour incessantes du fichier de la paie par les administrations sectorielles et les services utilisateurs des fonctionnaires; (iii) le problème relatif à l'identification et au paiement des inactifs, des veuves, des orphelins, des autres rentiers; (iv) le retard récurrent dans la transmission des rapports de paie par les banques intervenantes.

D'autres contraintes doivent faire l'objet de mesures appropriées. Il s'agit notamment de: (i) la faible culture bancaire des bénéficiaires; (ii) la méfiance du système bancaire par les bénéficiaires; (iii) la carence de banques dans les territoires d'accès difficile; (iv) la difficulté d'identifier, pour les banques intervenantes, les agents et fonctionnaires ainsi que les militaires ayant perdu leurs pièces d'identité. Pour rappel, cette réforme n'aurait pu être possible sans l'entreprenariat institutionnel, facteur déclencheur d'initiatives et générateurs d'opportunités. En effet, l'entrepreneuriat institutionnel met en jeu une véritable corrélation entre « offre et demande » de changements institutionnels (cf. Hederer 2010). La bancarisation de la paie des agents publics est une parfaite illustration de cette corrélation. En 2011, le gouvernement a formulé une offre de réforme institutionnelle visant à organiser le paiement des salaires des fonctionnaires sur des comptes bancaires en vue de renforcer le contrôle des effectifs et la maitrise de la masse salariale.

Cette offre de réforme a rencontré une forte demande de réforme par les banques commerciales désireuses d'améliorer leur taux de bancarisation. Un protocole d'accord a été conclu en date du 1er décembre 2012 entre le gouvernement et l'Association congolaise des banques pour l'ouverture de comptes bancaires des fonctionnaires moyennant couverture des frais de tenue des comptes par l'État. Quinze banques commerciales ont activement participé à la mise en œuvre effective de cette réforme.

Par ailleurs, l'opérationnalisation de la paie des salaires des fonctionnaires par la voie bancaire a rencontré la demande persistante des banques commerciales pour l'amélioration de l'inclusion financière ainsi que des ménages pour la sécurisation de leurs revenus.

Ainsi, comme illustré à la section précédente, nous pouvons retenir que les réformes impulsées par l'entrepreneuriat public créent des opportunités pour l'entrepreneuriat économique, et le cercle des réformes et de la croissance s'entretient par davantage d'entrepreneuriat économique générant des possibilités additionnelles de réformes institutionnelles pour des rendements croissants[15]. Dans le cas de la bancarisation de la paie des agents publics en RDC, les besoins de mise en œuvre de cette réforme dans les territoires reculés et non desservis par les banques ont généré un nouvel entrepreneuriat entre le gouvernement, les banques et les entreprises du secteur de télécommunication pour le recours au *mobile banking*.

**La réforme de la gestion des ressources humaines de l'administration publique**. Au regard des ambitions de la RDC visant à rejoindre, dans un délai proche, les économies des pays à revenu intermédiaire, et préparer ainsi les conditions de l'émergence, le gouvernement de la RDC s'est assigné comme objectif majeur de son programme d'action 2012-2016 approuvé par l'Assemblée nationale: « poursuivre et finaliser les réformes institutionnelles en vue de renforcer l'efficacité de l'État ».

La réforme de l'administration publique a ainsi été identifiée parmi les réformes institutionnelles indispensables à la réhabilitation des capacités d'action de l'État. Car, durant plusieurs années, le secteur public a accusé des dysfonctionnements (cadre juridique obsolète, absence d'un système efficace de gestion des ressources humaines du secteur public, service public vieillissant et grille des salaires de moins attrayants) se traduisant par des insuffisances empêchant l'administration publique d'être effectivement opérationnelle et de remplir son rôle moteur pour le fonctionnement harmonieux de l'appareil étatique.

---

[15] Cf. Henrekson et Sanandaji (2011).

Tableau 4: Aperçu des mesures structurelles mises en œuvre pour la modernisation de l'administration publique

| Priorités Sectorielles | Résultats 2016 |
|---|---|
| Redynamisation et coordination du pilotage de la réforme de l'administration publique | — Décret no. 12/028 du 03 août 2012 portant création, organisation et fonctionnement du CPMAP (Comité de Pilotage de la Modernisation de l'Administration Publique). |
| Modernisation du cadre juridique de la fonction publique | — Loi organique no. 16/001 du 03 mai 2016 fixant l'organisation et le fonctionnement des Services publics du Pouvoir central, des Provinces et ETD promulguée par le Chef de l'État.<br>— Loi no. 16/013 du 15 juillet 2016 portant statuts du personnel de carrière des services publics de l'État promulguée par le Chef de l'État. |
| Maitrise des effectifs et de la masse salariale de la fonction publique | — Bancarisation (à fin 2015): 987.314 agents de l'État sur 1.037.834 rémunérés ont été bancarisés (Fonction publique, PNC, FARDC, EPSP…).<br>— Data Center - SIGRH-Paie installé au ministère de la Fonction Publique: en attente d'être opérationnel avec l'interconnexion en cours des ministères (Budget, Finances, BCC). |
| Rajeunissement et renforcement des capacités de l'administration | — Décret no. 13/013 du 16 avril 2013 portant création de l'ENA.<br>— Trois vagues de 54, 107 et 543 jeunes universitaires recrutés par le BCECO sur concours (2010, 2011 et 2012) &admis sous statuts.<br>— Trois promotions de 60, 100 &100 jeunes énarques recrutés sur concours par la Fonction publique et formés à l'ENA (2014, 2015 &2016) &admis sous statuts. |
| Mise en œuvre de nouveaux cadres organiques: Directions standards (DEP, DRH, DAF, DNTIC) | — Décret no. 15/043 du 28 décembre 2015 portant fixation du cadre organique des structures à compétences horizontales communes à toutes les administrations centrales des ministères. |
| Mise en place de la caisse de retraite des fonctionnaires | — Décret no. 15/031 du 14 décembre 2015 portant organisation et fonctionnement de la Caisse nationale de sécurité sociale des agents publics de l'État (CNSSAP). |

En vue de relancer et assurer une meilleure coordination du pilotage de la réforme de l'administration publique, le gouvernement a mis en place, par voie du Décret no. 12/028 du 03 août 2012, le comité de pilotage pour la modernisation de l'administration publique (CPMAP). Ce dernier fut placé sous la direction du Premier ministre. Il avait pour ambition d'assurer les contacts réguliers entre les institutions et ministères clés dans la mise en œuvre de la réforme de l'administration publique. Le pilotage technique du ministère de la fonction publique a été renforcé par la création de la cellule de mise en œuvre de la réforme de l'administration publique par voie d'arrêté ministériel no. CAB.MIN/FP/J-CK/GELB/FMD/GMK/082/2012 du 12 décembre 2012. Grâce à cette dynamique, les institutions et ministères membres du CPMAP réunis sous la direction du premier ministre à savoir: le représentant de la Présidence de la République, les ministres de la fonction publique, du budget, des finances, du plan et de la décentralisation ont pu mettre à la disposition du gouvernement un cadre stratégique général pour la réforme et la modernisation de l'administration publique (CSRAP) approuvé en Conseil des ministres et adopté par les parties prenantes.

Enfin, en vue d'améliorer les prestations des services publics en faveur des usagers et démontrer dans les faits l'impact de la réforme et modernisation de l'administration publique, le gouvernement a lancé la mise en œuvre de centres d'excellence tels les guichets uniques. Toutefois, le cadre légal et institutionnel appelé à porter ces réformes structurelles est apparu suranné et peu adapté. Cette difficulté a donc amené le gouvernement à engager un dialogue institutionnel soutenu avec les institutions impliquées dans l'élaboration et la promulgation des lois, à savoir le Parlement et le Président de la République.

La modification du cadre légal et institutionnel initiée par le gouvernement a donc visé à revisiter le cadre organisationnel de l'administration publique et à tourner celle-ci vers la prestation de service public, conformément à la Charte africaine sur les valeurs et les principes du service public. Ainsi, ont été élaborées et promulguées les deux lois majeures suivantes de modernisation du cadre juridique de la fonction publique: (i) la Loi organique no. 16/001 du 03 mai 2016 fixant l'organisation et le fonctionnement des services publics du pouvoir central, des provinces et ETD; (ii) la Loi no. 16/013 du 15 juillet 2016 portant statuts du personnel de carrière des services publics.

Une innovation importante pour la transparence et la gouvernance est l'introduction de principes régissant les rapports entre Administration et usagers dont entre autres la qualité et l'efficience des services rendus, l'évaluation des services par le public impliquant la société civile ainsi que la célérité et le respect des délais de traitement. Quant au statut du personnel, il a permis de relever l'âge de la retraite relevé de 55 à 65 ans, d'instaurer un régime contributif de la pension de retraite ainsi que des organes administratifs consultatifs paritaires en vue de garantir la transparence dans la gestion disciplinaire du personnel. Enfin, le gouvernement a soumis à l'examen du Parlement le projet de loi fixant les règles sur la sécurité sociale des agents publics de l'État.

D'autres textes règlementaires structurant ont été adoptés dont: (i) le Décret no. 15/031 du 14 décembre 2015 portant création, organisation et fonctionnement d'un établissement public dénommé Caisse nationale de sécurité sociale des agents publics de l'État (CNSSAP); (ii) le Décret no. 15/043 du 28 décembre 2015 portant fixation du cadre organique des structures à compétences horizontales communes à toutes les administrations centrales des ministères.

Sur le plan opérationnel, plusieurs mesures structurelles ont été mises en œuvre en vue d'améliorer la gestion qualitative des ressources humaines du secteur public pour un meilleur rendement. Il y a lieu de citer:

(i) La *budgétisation préalable des besoins en ressources humaines*. Il s'agit en effet d'une exigence de la loi no. 11/011 du 13 juillet 2011 relatives aux finances publiques. Aussi, l'annexe relative à l'état des plafonds des autorisations d'emplois rémunérés par l'État ainsi qu'à la création d'emplois nouveaux doit faire partie des projets de loi de finances à soumettre à l'Autorité budgétaire pour sanction.

(ii) L'*organisation, par un service dédié compétent, des recrutements transparents et compétitifs conformément aux emplois nouveaux budgétisés.* La création de l'École nationale de l'administration (ENA) pour la formation initiale des jeunes cadres universitaires et la formation continue des cadres et agents de carrière en est la pierre angulaire. Depuis sa création en 2014, l'ENA a recruté sur concours et formé 60 jeunes universitaires la première année et 635 jeunes la deuxième année, dont 100 Énarques. Ces jeunes universitaires sont tous destinés au rajeunissement des administrations publiques.

(iii) La *mise en place du système intégré de gestion des ressources humaines et de la paie pour la maitrise des effectifs et de la masse salariale.* Le paramétrage dudit système informatisé a été rendu possible grâce à la finalisation du recensement biométrique et la bancarisation de la paie des agents et fonctionnaires de l'État. La finalisation de cette mesure interviendra avec l'étape d'interconnexion entre les ministères de la Fonction Publique, du Budget, des Finances et de la Banque centrale.

(iv) La *mise en œuvre d'une gestion prévisionnelle des emplois et des compétences (GPEC) ainsi que d'une démarche de la performance par de véritables directions des ressources humaines*. Les exigences de gouvernance pour l'efficacité de l'administration, imposent d'organiser de manière rigoureuse l'adéquation « profil-poste ». En vue d'instaurer cette démarche de rendement, les administrations ministérielles ont été dotées de cadres organiques revisités qui intègrent les « fonctions-support » majeures pour leur fonctionnement et la mise en œuvre de leurs missions. Les ministères pilotes ont ainsi été dotés de nouvelles directions standards (Gestion des ressources humaines, Administration et finances, Études et Planification, Nouvelles technologies de l'information et de la communication). La direction en charge des ressources humaines devra implémenter la gestion axée sur les résultats à travers les outils de notation annuelle.

Quant à la politique salariale, le gouvernement s'est engagé dans une démarche visant à tenir compte des recettes internes mobilisées, en veillant à ne pas accroître les recrutements sans tenir compte des prévisions budgétaires. Il s'agit ainsi de procéder à une rationalisation de la politique salariale à travers un cadre budgétaire à moyen terme défini sur base du cadrage macroéconomique indiquant le plafond de la masse salariale. La création de la CNSSAP visait à mettre en œuvre une véritable politique de prévoyance sociale assise sur une caisse de retraite organisée selon des mécanismes rigoureux de gestion des risques fiduciaires. Par cette réforme, le gouvernement entend migrer d'un régime de retraite actuellement octroyé à un régime contributif qui permettra de soulager le Trésor public et d'assurer une planification efficace des départs à la retraite.

Le gouvernement a également engagé un dialogue institutionnel en vue de la mise en place d'un mécanisme de contrôle du fonctionnement des services publics. Ce mécanisme devrait se charger du suivi et de l'évaluation des administrations publiques dans la mise en œuvre des procédures administratives tel que prescrit par la loi.

Enfin, comme le préconise la stratégie du gouvernement, ce dernier s'est lancé dans la mise en œuvre de guichets uniques pour l'amélioration rapide de la qualité des services publics en faveur des usagers. Ainsi, le gouvernement a successivement pris le Décret no. 12/045 du 1er novembre 2012 portant Création du guichet unique de création des entreprises modifié par le décret no. 14/014 du 08 mai 2014 et le Décret no. 15/019 du 15 octobre 2015 instituant un Guichet unique du commerce extérieur.

**Autres réformes**. Plusieurs autres réformes techniques sectorielles ont été inscrites dès le départ sous le PAG 2012-2016. Au nombre de ces réformes initiées par le gouvernement, l'on peut citer: (i) la mise en œuvre de la TVA; (ii) l'adhésion au Traité Organisation pour l'Harmonisation du Droit des Affaires en Afrique, OHADA; (iii) la rationalisation des procédures pour la création d'entreprises et donc l'amélioration du climat des affaires par la mise en place d'un guichet unique; (iv) la publication régulière des ressources mobilisées par les régies financières dans le secteur des ressources naturelles à travers l'ITIE (Initiative pour la Transparence dans les Industries Extractives); (v) l'entrée en vigueur de la convention de New York sur l'exécution des sentences arbitrales rendues à l'étranger; (vi) la réforme des entreprises publiques; (vii) l'ouverture à la concurrence de plusieurs secteurs d'activités par l'adoption des textes des lois notamment: la loi sur l'agriculture, le code minier, la loi sur l'électricité, le code des assurances, la loi sur le crédit-bail, la loi sur les hydrocarbures, l'installation progressive des tribunaux de commerce, l'opérationnalisation du registre de commerce et du crédit mobilier; (viii) la création de la Cellule d'Analyse des Indicateurs de Développement (CAID) dans l'objectif de mettre l'information à la disposition de la population tel que repris dans une partie du modèle.

# 6. Conclusion

Ce papier utilise le cadre de la théorie des jeux coopératifs pour examiner les interactions entre le leadership et les changements institutionnels. Nous réaménageons ce cadre d'analyse, d'une part, pour établir analytiquement une différence entre un dirigeant politique et un leader politique et, d'autre part, pour étudier formellement l'incidence des interactions mutuelles entre un dirigeant politique et ses partisans sur la conduite des réformes structurelles.

À la différence d'un dirigeant politique, nous définissons un leader politique comme une personnalité qui a le pouvoir d'inspirer les changements, mais aussi et surtout de les imprimer effectivement au gré des groupes de pression (souvent, une minorité) ou obstacles divers. De ce qui précède, le leadership peut donc être analytiquement perçu comme la probabilité qu'un dirigeant politique soit susceptible d'inspirer et de provoquer effectivement un changement en faveur de la majorité de la population. Plus important encore, nous montrons formellement qu'un dirigeant politique peut être à la fois partisan et non-partisan, alors qu'un leader politique ne peut qu'être non-partisan.

Suivant cette distinction, nous caractérisons la probabilité de réussite d'un changement institutionnel, ainsi que les gains éventuels de ce changement pour la population bénéficiaire. En nous basant sur ce cadre d'analyse standard simple, nous soutenons que si dans les pays en développement, les changements institutionnels ne profitent qu'à la minorité de la population, c'est parce que fondamentalement les dirigeants politiques sont pour la plupart de cas « partisans », c'est-à-dire ils ne sont pas véritablement des « leaders ». En parallèle, nous montrons à l'aide des quelques exemples pertinents, tirés de l'expérience congolaise entre 2010 et 2016, que les changements institutionnels peuvent effectivement profiter à la majorité de la population, lorsque les dirigeants politiques sont véritablement non-partisans, i.e. lorsqu'ils font preuve d'un leadership efficient dans la conduite des réformes institutionnelles. Parmi les exemples innovateurs ayant marqué la gouvernance publique entre 2010 et 2016 et ayant requis un leadership efficient, ont été principalement considérées: la bancarisation de la paie des agents et fonctionnaires de l'État; la modernisation de la gestion des ressources humaines de l'administration publique; et l'implémentation effective des lettres de mission.

## Bibliographie